\documentstyle[aps]{revtex}
\begin{document}
\author{Guoying Chee}
\address{China Center of Advanced Science and Technology (World Laboratory)\\
P. O. Box 8730, Beijing 100080, China\\
Department of Physics, Normal University of Liaoning , Dalian 116029, China\\
Email address: gyq@mail.dlptt.ln.cn}
\author{Yanhua Jia}
\address{Department of Applied Mechanics, Beijing Institute of Technology, Beijing\\
100081, China }
\title{Self-Dual Conformal Supergravity and the Hamiltonian Formulation }
\maketitle

\begin{abstract}
In terms of Dirac matrices the self-dual and anti-self-dual decomposition of
a conformal supergravity is given and a self-dual conformal supergravity
theory is developed as a connection dynamic theory in which the basic
dynamic variables include the self-dual spin connection {\it i.e. }the
Ashtekar connection rather than the triad. The Hamiltonian formulation and
the constraints are obtained by using the Dirac-Bergmann algorithm.

\end{abstract}

\section{{}Introduction}

Among the various approaches to the construction of a unified model for the
fundamental interactions including gravity many attempts have been made to
write down gravity as a Yang-mills type gauge theory where the basic
dynamical object is a connection one-form associated with some group. In
this approach the metric (the tetrad) and the Lorentz connection are
identified as different components of a connection one-form. A famous
example is the MacDowell-Mansouri gravitational formalism [1] which mimics,
as much as possible, the Yang-Mills type gauge theory in four space-time
dimensions and has been successfully applied to construct different
supergravity theories [2].

In 1986, a somewhat different, but nonetheless related approach was
initiated [3] with introducing the new variables in general relativity which
can be thought of as a Yang-Mills connection one form on a spacelike
hypersurface. much of the success associated with the new variables appears
to be intimately related to their character as gauge fields. Not long after
Ashtekar's results, Jacobson was able to formulate supergravity in the new
variables [4]. Further, Capovilla, Jacobson, and dell developed a
pure-connection theory of gravity, {\it i.e., }a formulation of general
relativity without metric [5]. On the other hand, as early as 1974, Chern
and Simons constructed a pure-connection theory of gravity [6].

Recently, several authors [7-11] proposed a self-dual generalization of the
MacDowell-Mansouri formalism which includes the Ashtekar-Jacobson theory as
well as Yang-Mills theory starting from the (anti-) de Sitter group. Beside
the de Sitter or Poincare supergravity there is another class of
supergravity {\it i. e., }the conformal supergravity.. And it is conformal
supergravity that provides a true unification of gravity and gauge fields.
By gauging SU(2,2%
%TCIMACRO{\TEXTsymbol{\vert}}
%BeginExpansion
\mbox{$\vert$}%
%EndExpansion
1) group and imposing some constraints on curvature a simple conformal
supergravity has been developed by Nieuwenhuizen {\it et al }[2, 12-14].
However, in this theory tetrad rather than connection was taken to be a
basic dynamical variable in the second-order formalism. Therefore, it is not
a connection dynamical but a geomitrodynamical theory in a sense. On the
other hand , the Lagrangian in this theory is quadratic in the curvature and
then is different from the Einstein-Hilbert Lagrangian.. It is reasonable to
expect that one of the basic dynamical variable be the connection instead of
the tetrad. In this paper we show that this is the case. A self-dual
conformal supergravity is developed and its Hamiltonian formulation is
obtained. In Sec. 2 we start by recalling the conformal superalgebra su(2,2%
%TCIMACRO{\TEXTsymbol{\vert}}
%BeginExpansion
\mbox{$\vert$}%
%EndExpansion
1) and then define the dual of a element of su(2,2%
%TCIMACRO{\TEXTsymbol{\vert}}
%BeginExpansion
\mbox{$\vert$}%
%EndExpansion
1), its self-dual and anti-self-dual part using the Dirac matrix $\gamma _5.$
The Lagrangian of the conformal supergravity is constructed in Sec. 3 and
then the decomposition into self-dual and anti-self-dual is given in Sec. 4,
a self-dual conformal supergravity is obtained. In Sec. 5 its Hamiltonian
formulation is investigated and the structure of the constraints is
discussed. In the appendix we list the Poison brackets of the constraints.
The complicated structure of these Poison brackets makes the classification
of the constraints impossible. The Dirac brackets, however, permit us to get
rid of the second class constraints. We obtain a constrained Hamiltonian
system. The action is first order in the time derivatives and the
Hamiltonian results to be a linear combination of the constraints.

\section{The conformal superalgebra su(2,2$|$1)}

The conformal superalgebra su(2,2$|$1) is given by [7]

$\left[ M_{IJ},M_{KL}\right] =\eta _{JK}M_{IL}+\eta _{IL}M_{JK}-\eta
_{JL}M_{IK}-\eta _{IK}M_{JL}$

$\left[ M_{IJ},P_K\right] =\eta _{JK}P_I-\eta _{IK}P_J,\;\left[
M_{IJ},K_K\right] =\eta _{JK}K_I-\eta _{IK}K_J,$

$\left[ P_I,D\right] =P_I,\;\left[ K_I,D\right] =-K_I,\;\left[
P_I,K_J\right] =2\eta _{IJ}D-2M_{IJ},$

$[Q^\alpha ,M_{IJ}]=\frac 12(\gamma _{IJ})^\alpha \ _\beta \;Q^\beta
,\;\left[ S^\alpha ,M_{IJ}\right] =\frac 12(\gamma _{IJ})^\alpha \ _\beta
\;S^\beta ,$

$\left[ S^\alpha ,P_I\right] =(\gamma _I)^\alpha \ _\beta \;Q^\beta
,\;[Q^\alpha ,K_I]=-(\gamma _I)^\alpha \ _\beta \;S^\beta ,$

$[Q^\alpha ,D]=\frac 12Q^\alpha ,\;\left[ S^\alpha ,D\right] =-\frac 12%
S^\alpha ,$

$[Q^\alpha ,A]=-\frac 34(\gamma _5)^\alpha \ _\beta \;Q^\beta ,\;\left[
S^\alpha ,A\right] =\frac 34(\gamma _5)^\alpha \ _\beta \;S^\beta ,$

$\left\{ Q^\alpha ,Q^\beta \right\} =-\frac 12(\gamma ^IC^{-1})^{\alpha
\beta }P_I,\;\left\{ S^\alpha ,S^\beta \right\} =\frac 12(\gamma
^IC^{-1})^{\alpha \beta }K_I,$

$\left\{ Q^\alpha ,S^\beta \right\} =-\frac 12(C^{-1})^{\alpha \beta }D+%
\frac 12(\gamma ^{IJ}C^{-1})^{\alpha \beta }M_{IJ}+(\gamma _5C^{-1})^{\alpha
\beta }A,$%
\begin{equation}  \label{1}
\end{equation}
where

\begin{equation}
\left\{ \gamma _I,\gamma _J\right\} =2\eta _{IJ},\left\{ \gamma _I,\gamma
_5\right\} =0,\;\gamma _5^2=-1,
\end{equation}
and

\begin{equation}
\gamma ^{IJ}=\frac 12(\gamma ^I\gamma ^J-\gamma ^J\gamma ^I).
\end{equation}
To fulfill these relations we can choose the matrix representations of the
Bose basis

\begin{eqnarray}
P_I &=&-\frac 12\gamma _I(1+i\gamma _5),\;K_I=\frac 12\gamma _I(1-i\gamma
_5),  \nonumber \\
M_{IJ} &=&\frac 12\gamma _{IJ},\;D=\frac i2\gamma _5,\;A=-\frac i4I,
\end{eqnarray}
and the Majorana spinor representations of the Fermi basis

\begin{equation}
Q^\alpha =\left( 
\begin{array}{c}
Q^A \\ 
\overline{Q}_{A^{\prime }}
\end{array}
\right) ,\;and\;S^\alpha =\left( 
\begin{array}{c}
S^A \\ 
\overline{S}_{A^{\prime }}
\end{array}
\right) .
\end{equation}
In this paper we adopt the following index notation: $I,J,K,L,\ldots $ are
group indices; $\alpha ,\beta ,\ldots $ are Majorana spinor indices; $\mu
,\nu ,\rho ,\ldots $ are spacetime indices ; $i,j,k,\ldots $ are spatial
indices, $A,B,\ldots $ and $A^{\prime },B^{\prime },\ldots $ are used to
denote $SL(2C)$ spinor indices.

Using the identities

\begin{equation}
\epsilon ^{IJKL}\gamma _{IJ}=2\gamma _5\gamma ^{KL},\;\;and\epsilon
^{IJKL}\gamma _I\gamma _J\gamma _K=-6\gamma _5\gamma ^L,
\end{equation}
for any element $O$ of SU(2,2$|$1), we can define its dual by

\begin{equation}
O*\;=\gamma _5O
\end{equation}
then the self-dual and the anti-self-dual parts of $O$ are given by,
respectively,

\begin{eqnarray*}
O^{+} &=&\frac 12(O-iO*)=\frac 12(1-i\gamma _5)O, \\
O^{-} &=&\frac 12(O+iO*)=\frac 12(1+i\gamma _5)O.
\end{eqnarray*}

\section{Conformal supergravity}

Introducing the su(2,2$|$1) algebra valued connection one-form

\begin{eqnarray}
\Gamma &=&\omega +{\bf e}+{\bf f+b+A+}\psi {\bf +}\phi  \nonumber \\
\ \ &=&\frac 12\omega ^{IJ}\otimes M_{IJ}+e^I\otimes P_I+f^I\otimes
K_I+b\otimes D+  \nonumber \\
&&a\otimes A+\psi ^\alpha \otimes Q_\alpha +\phi ^\alpha \otimes S_\alpha ,
\end{eqnarray}
and its curvature

\begin{equation}
\Omega =D\Gamma =d\Gamma +\frac 12[\Gamma ,\Gamma ],
\end{equation}
we can compute

\begin{eqnarray}
\Omega &=&\Omega (M)+\Omega (P)+\Omega (K)+\Omega (D)+\Omega (A)+\Omega
(Q)+\Omega (S)  \nonumber \\
\ &=&\frac 12\Omega ^{IJ}(M)\otimes M_{IJ}+\Omega ^I(P)\otimes P_I+\Omega
^I(K)\otimes K_I+\Omega (D)\otimes D  \nonumber \\
&&+\Omega (A)\otimes A+\Omega ^\alpha (Q)\otimes Q_\alpha +\Omega ^\alpha
(S)\otimes S_\alpha ,
\end{eqnarray}
\ where

$\Omega ^{IJ}(M)=d\omega ^{IJ}+\omega ^{IK}\bigwedge \omega
_K{}^J-4(e^I\bigwedge f^J-e^J\bigwedge f^I)+\frac 12\overline{\psi }\gamma
^{IJ}\phi ,$

$\Omega ^I(P)=de^I+\omega ^{IJ}\bigwedge e_J-\frac 1{4\sqrt{2}}\overline{%
\psi }\bigwedge \gamma ^I\psi -2e^{I\bigwedge }b,$

$\Omega ^I(K)=df^I+\omega ^{IJ}\bigwedge f_J+\frac 1{4\sqrt{2}}\overline{%
\phi }\bigwedge \gamma ^I\phi +2f^{I\bigwedge }b,$

$\Omega $$(D)=db-2e^I\bigwedge f_I+\frac 14\overline{\psi }\bigwedge \phi ,$

$\Omega (A)=da+\frac 14\overline{\psi }\bigwedge \gamma _5\phi $

$\Omega (Q)=d\psi +\frac 14\omega ^{IJ}\bigwedge \gamma _{IJ}\psi
+b\bigwedge \psi +3a\bigwedge \gamma _5\psi +\sqrt{2}\gamma _Ie^I\bigwedge
\phi ,$

$\Omega (S)=d\phi +\frac 14\omega ^{IJ}\bigwedge \gamma _{IJ}\phi
-b\bigwedge \phi -3a\bigwedge \gamma _5\phi -\sqrt{2}\gamma _Jf^I\bigwedge
\psi .$%
\begin{equation}  \label{12}
\end{equation}

For a gauge theory its Lagrangian can be chosen among the four types

\[
\left\langle \Omega \bigwedge \Omega *\right\rangle ,\left\langle \Omega
\bigwedge \Omega \right\rangle ,\left\langle *\Omega \bigwedge \Omega
\right\rangle ,and\;\left\langle *\Omega \bigwedge \Omega *\right\rangle 
\]
$,$where $*\Omega \;$denotes the usual Hodge dual of $\Omega \;$with respect
to the spacetime metric and $\left\langle ,\right\rangle \;$is the Killing
inner product defined in the superalgebra su(2,2$|$1). In the bosonic sector

\[
\left\langle O,O^{\prime }\right\rangle =Tr(OO^{\prime }), 
\]
and in the fermionic sector

\[
\left\langle O,O^{\prime }\right\rangle =\overline{O}O^{\prime }, 
\]
where $\overline{O}\;$is the Dirac conjugation of $O.\;$Using (11) and (12)
we can compute, for example,

\begin{equation}
\left\langle \Omega \bigwedge \Omega *\right\rangle =\left\langle \Omega
(M)\bigwedge \Omega (M)*\right\rangle +\left\langle \Omega (D)\bigwedge
\Omega (A)*\right\rangle +\left\langle \Omega (A)\bigwedge \Omega
(D)*\right\rangle ,
\end{equation}
where

\begin{eqnarray}
\left\langle \Omega (M)\bigwedge \Omega (M)*\right\rangle &=&-\frac 14%
\epsilon ^{IJKL}\left\langle R(\omega )_{IJ}\bigwedge R(\omega
)_{KL}\right\rangle -8\epsilon ^{IJKL}e_I\bigwedge f_J\bigwedge e_K\bigwedge
f_L  \nonumber \\
&&\ -\frac 1{16}\epsilon ^{IJKL}\overline{\psi }\gamma _{IJ}\phi \bigwedge 
\overline{\psi }\gamma _{KL}\phi ,
\end{eqnarray}
with $R(\omega )_{IJ}=D\omega _{IJ}=d\omega _{IJ}+\omega
\!_I^{\;\;\;\;K}\bigwedge \omega _{KJ}{}$ and

\begin{eqnarray}
&&\left\langle \Omega (D)\bigwedge \Omega (A)*\right\rangle +\left\langle
\Omega (A)\bigwedge \Omega (D)*\right\rangle  \nonumber \\
&=&2da\bigwedge db+\frac i2(\psi _A\bigwedge \phi ^A+\overline{\psi }%
^{A^{\prime }}\bigwedge \overline{\phi }_{A^{\prime }})\bigwedge db 
\nonumber \\
&&-da\bigwedge (4e^I\bigwedge f_I-\frac 12\psi _A\bigwedge \phi ^A+\frac 12%
\overline{\psi }^{A^{\prime }}\bigwedge \overline{\phi }_{A^{\prime }}) 
\nonumber \\
&&-ie^I\bigwedge f_I\bigwedge (\psi _A\bigwedge \phi ^A+\overline{\psi }%
^{A^{\prime }}\bigwedge \overline{\phi }_{A^{\prime }})\ \   \nonumber \\
&&+\frac i8(\psi _A\bigwedge \phi ^A\bigwedge \psi _B\bigwedge \phi ^B-%
\overline{\psi }^{A^{\prime }}\bigwedge \overline{\phi }_{A^{\prime
}}\bigwedge \overline{\psi }^{B^{\prime }}\bigwedge \overline{\phi }%
_{B^{\prime }}).
\end{eqnarray}

It is notable that the property

\begin{equation}
\overline{\psi }\phi =-\overline{\phi }\psi
\end{equation}
leads to

\begin{equation}
\left\langle \Omega (Q)\bigwedge \Omega (S)*\right\rangle =-\left\langle
\Omega (S)\bigwedge \Omega (Q)*\right\rangle
\end{equation}
and then there are no dynamical terms of the Fermi fields $\psi \;$and $\phi
\;$in the Lagrangian, which is different from the Lagrangian given by
Nieuwenhuizen [6--8]:

\begin{equation}
{\cal L}=4\left\langle \Omega (M)\bigwedge \Omega (M)*\right\rangle
-32\left\langle \Omega (D)\bigwedge \Omega (A)*\right\rangle +\left\langle
\Omega (Q)\bigwedge \Omega (S)*\right\rangle
\end{equation}
It is notable that the Lagragian (13) is obtained without using the
constraints on curvature which is indispensable for the Nieuwenhuizen
approach.

\section{Self-dual conformal supergravity}

Using the definition of the self-dual and the anti-self-dual introduced in
Sec. 2, the connection $\Gamma \;$can be decomposed into two parts:

\[
\;\Gamma =\Gamma ^{+}+\Gamma ^{-},\ 
\]
where

\begin{equation}
\Gamma ^{\pm }=\omega ^{\pm }+{\bf e^{\pm }}+{\bf f^{\pm }+b^{\pm }+A^{\pm }+%
}\psi {\bf ^{\pm }+}\phi ^{\pm },
\end{equation}
and so can the curvature

\begin{equation}
\Omega =\Omega ^{+}+\Omega ^{-},
\end{equation}
where

\ 
\begin{equation}
\Omega ^{\pm }=\Omega ^{\pm }(M)+\Omega ^{\pm }(P)+\Omega ^{\pm }(K)+\Omega
^{\pm }(D)+\Omega ^{\pm }(A)+\Omega ^{\pm }(Q)+\Omega ^{\pm }(S).
\end{equation}

\ 

Since

\begin{equation}
\left\langle \Omega \bigwedge \Omega *\right\rangle =\left\langle \Omega
\bigwedge \Omega *\right\rangle ^{+}+\left\langle \Omega \bigwedge \Omega
*\right\rangle ^{-},
\end{equation}
where

\begin{eqnarray}
\left\langle \Omega \bigwedge \Omega *\right\rangle ^{+} &=&\left\langle
\Omega ^{+}\bigwedge \Omega *^{+}\right\rangle =i\left\langle \Omega
^{+}\bigwedge \Omega ^{+}\right\rangle ,  \nonumber \\
\left\langle \Omega \bigwedge \Omega *\right\rangle ^{-} &=&\left\langle
\Omega ^{-}\bigwedge \Omega *^{-}\right\rangle =-i\left\langle \Omega
^{-}\bigwedge \Omega ^{-}\right\rangle ,
\end{eqnarray}
and $\left\langle \Omega \bigwedge \Omega *\right\rangle \;$does not include
dynamical terms of the fields $\psi $\ and\ $\phi $, we choose the self-dual
part of the Nieuwenhuizen$\;$Lagrangian

\begin{equation}
{\cal L}=4\left\langle \Omega (M)\bigwedge \Omega (M)*\right\rangle
^{+}-32\left\langle \Omega (D)\bigwedge \Omega (A)*\right\rangle
^{+}+8\left\langle \overline{\Omega (Q)}\bigwedge \Omega (S)*\right\rangle
\end{equation}
to be the Lagrangian of the self-dual conformal supergravity theory instead
of\ $\left\langle \Omega \bigwedge \Omega *\right\rangle $.

In order to obtain the explicit expression of$\;{\it L}\;$we use the matrix
representation of the superalgebra su(2,2$|$1). In the chiral representation
of the Dirac matrices we have

\begin{eqnarray}
\gamma ^I &=&\sqrt{2}\left[ \; 
\begin{array}{cc}
0 & \sigma ^{IAA^{\prime }} \\ 
(\sigma ^I{}_{AA^{\prime }})^t & 0
\end{array}
\right] ,  \nonumber \\
\gamma ^{IJ} &=&\left[ 
\begin{array}{cc}
\sigma ^{IAA^{\prime }}\sigma ^J{}_{BA^{\prime }}-\sigma ^{JAA^{\prime
}}\sigma ^I{}_{BA^{\prime }} & 0 \\ 
0 & \sigma ^I{}_{AA^{\prime }}\sigma ^{JAB^{\prime }}-\sigma
^J{}_{AA^{\prime }}\sigma ^{IAB^{\prime }}
\end{array}
\right]  \nonumber \\
\; &=&\left[ 
\begin{array}{cc}
\gamma ^{+IJA}{}_B & 0 \\ 
0 & \gamma ^{-IJ}{}_{A^{\prime }}{}^{B^{\prime }}
\end{array}
\right] ,  \nonumber \\
A &=&-\frac i4\;\left[ 
\begin{array}{cc}
I & 0 \\ 
0 & I
\end{array}
\right] ,and\;D=-\frac 12\;\left[ 
\begin{array}{cc}
I & 0 \\ 
0 & -I
\end{array}
\right] .
\end{eqnarray}

In this representation the spin connection$\;\omega \;$and its curvature$%
\;R(\omega )=d\omega +\frac 12\;[\omega ,\omega ]\;$have the two component
spinor forms

\[
\omega =\left[ 
\begin{array}{cc}
\omega ^{+A{}}{}_B & 0 \\ 
0 & \omega ^{-}{}_{A^{\prime }}{}^{B^{\prime }}
\end{array}
\right] , 
\]
and

\begin{equation}
R(\omega )=\left[ 
\begin{array}{cc}
R^{+}(\omega )^A\!_B & 0 \\ 
& R^{-}(\omega )_{A^{\prime }}\!^{B^{\prime }}
\end{array}
\right]
\end{equation}
\ where$\;\omega $$^{+A{}}{}_B=\frac 12\omega ^{IJ}\gamma _{IJ}{}^A{}_B\;$and%
$\;R^{+A{}}{}_B(\omega )=\frac 12R(\omega )^{IJ}\gamma _{IJ}{}^A{}_B\;$are
the self-dual parts of $\omega $\ and $R(\omega )$ respectively. From\ (25)
we get$\;\gamma ^{IJC}$\/$_D\gamma _{IJ\text{\/}}$\/$^A$\/$_B=4\epsilon
^{CA}\epsilon _{DB}-4\delta _B$\/$^C\delta _A$\/$^D,\;$and$\;\gamma ^{IJ}$\/$%
_{C^{\prime }}$\/$^{D^{\prime }}\gamma _{IJ\text{\/}}$\/$^A$\/$_B$ $=0.\;$%
Then we can obtain

$4\left\langle \Omega (M)\bigwedge \Omega (M)*\right\rangle ^{+}$

$=i\{4B\cdot B-4\overline{C}\cdot \overline{C}+32(e^{AA^{\prime
}}f_{BA^{\prime }}e^{BB^{\prime }}f_{AB^{\prime }}-f^{AA^{\prime
}}e_{BA^{\prime }}e^{BB^{\prime }}f_{AB^{\prime }})+2\psi ^A\phi _B\psi
^B\phi _A\}\sigma d^4x,$%
\begin{equation}  \label{27}
\end{equation}
where $\sigma =\det (\sigma _I{}^{AA^{\prime }}),\;f^{AA^{\prime
}}=f^I\sigma _I\,^{AA^{\prime }}\;$and

\begin{eqnarray}
B\cdot B &=&\frac 1{16}R_{AE^{\prime }B}{}^{E^{\prime }}{}_{CF^{\prime
}D}{}^{F^{\prime }}R^{AG^{\prime }B}{}_{G^{\prime }}{}^{CH^{\prime
}D}{}_{H^{\prime }},  \nonumber \\
\overline{C}\cdot \overline{C} &=&\frac 1{16}R_{EA^{\prime
}}{}^E{}_{B^{\prime }CF^{\prime }D}{}^{F^{\prime }}R^{GA^{\prime
}}{}_G{}^{B^{\prime }CH^{\prime }D}{}_{H^{\prime },}
\end{eqnarray}
and the spacetime indices $\mu ,\nu ,......\;$have been transformed to
spinor indices $AB^{\prime },\;CD^{\prime },......$ using the formula, for
example

\begin{equation}
V^{AB^{\prime }}=V^\mu e_\mu {}^I\sigma _I{}^{AB^{\prime }}.
\end{equation}

From the matrix expression of $D$\ and $A$\ we see that

\[
\Omega ^{+}(A)=\Omega ^{-}(A),\;\Omega ^{+}(D)=-\Omega ^{-}(D),\; 
\]
and then

\begin{equation}
\left\langle \Omega (D)\bigwedge \Omega (A)*\right\rangle ^{+}.=\left\langle
\Omega (D)\bigwedge \Omega (A)*\right\rangle ^{-}=\frac 12\left\langle
\Omega (D)\bigwedge \Omega (A)*\right\rangle .
\end{equation}
Using (12) and (25) and (31) we have

\begin{eqnarray}
&&-32\left\langle \Omega (D)\bigwedge \Omega (A)*\right\rangle ^{+} 
\nonumber \\
&=&-16da\bigwedge db-8i\psi _A\bigwedge \phi ^A\bigwedge db+da\bigwedge
(32e^I\bigwedge f_I-8\psi _A\bigwedge \phi ^A)+  \nonumber \\
&&16i\psi _A\bigwedge \phi ^A\bigwedge e^I\bigwedge f_I-2i\psi _A\bigwedge
\phi ^A\bigwedge \psi _B\bigwedge \phi ^B.
\end{eqnarray}
From (12), (11), (5), (20) and (8) we obtain

\begin{eqnarray}
&&8i\overline{\Omega (Q)}^{+}\bigwedge \Omega (S)^{+}  \nonumber \\
&=&8iD\psi _A\bigwedge D\phi ^A-8(3a-ib)\bigwedge (D\psi ^A\bigwedge \phi
_A-D\phi ^A\bigwedge \psi _A)+  \nonumber \\
&&16(3a-ib)\bigwedge (\psi ^A\bigwedge \;f_{AA^{\prime }}\bigwedge \overline{%
\psi }^{A^{\prime }}-\phi ^A\bigwedge \sigma _{AA^{\prime }}\bigwedge 
\overline{\phi }^{A^{\prime }})-  \nonumber \\
&&16i(D\psi ^A\bigwedge f_{AA^{\prime }}\bigwedge \overline{\psi }%
^{A^{\prime }}+D\phi ^A\bigwedge \sigma _{AA^{\prime }}\bigwedge \overline{%
\phi }^{A^{\prime }})-32i\overline{\phi }^{A^{\prime }}\sigma _{AA^{\prime
}}e^A\,_{B^{\prime }}\overline{\psi }^{B^{\prime }}.
\end{eqnarray}

Equation (24) with (27), (32) and (33) gives the Lagrangian for a self -dual
conformal supergravity.

\section{Hamiltonian formulation}

Following standard methods [15,16] a 3+1 decomposition of the Lagrangian can
be carried out to pass on to the Hamiltonian framework. In this
decomposition the tetrad variables $\sigma $\/$_\mu $\/$^{AA^{\prime }}\;$%
are split into $\sigma $\/$_0$\/$^{AA^{\prime }}\;$and$\;\sigma
\/_i\/^{AA^{\prime }}\;\;$( $i,\;j,\;\ldots \;$=\ 1,\ 2,\ 3). The spatial
spinor-valued forms$\;\sigma \/_i\/^{AA^{\prime }}\;$ determine the spatial
metric $q_{ij}=-tr\sigma _i\sigma _j\;$on a surface $\Sigma _t\;$with$\;t\;$%
= const. The spinor version $n^{AA^{\prime }}\;$of the unit timelike future
directed normal $n^\mu \;$to $\Sigma _t$\ can be used together with the $%
\sigma \/_i\/^{AA^{\prime }}\;$to make a basis for the space of spinors with
one unprimed and one primed index. It is determined by the $\sigma
\/_i\/^{AA^{\prime }}\;$through the conditions $n_{AA^{\prime }}$ $\sigma
\/_i\/^{AA^{\prime }}=0,\;n_{AA^{\prime }}$ $n^{AA^{\prime }}=-1.$ The
remaining variables$\;\sigma \/_0\/^{AA^{\prime }}$ can be expanded out as

\begin{equation}
\sigma \/_0\/^{AA^{\prime }}=\;Nn^{AA^{\prime }}+N^i\sigma
\/_i\/^{AA^{\prime }},
\end{equation}
where\ $N$ and $N^i\;$are the lapse and shift, respectively. Similarly the
other forms, {\it e. g. }the$\;\psi _\mu $ \/$^A$ are split in to$\;\psi
_0\/^A,\;$and$\;\psi _i\/^A\ $and their conjugates. Then a 3+1 decomposition
of the Lagrangian can be computed:

\begin{eqnarray}
{\cal L} &=&\;\widetilde{p}^i\/_A\/^B(\omega )\/\stackrel{.}{\omega }%
_i{}^A\/_B\;+\;\widetilde{p}^i(a)\stackrel{.}{a}_i+\;\widetilde{p}^i(b)%
\stackrel{.}{b}_i+\stackrel{.}{\psi }_i{}^A\/\;\widetilde{\pi }^i\/_A\/(q)+%
\stackrel{.}{\varphi }_i{}^A\/\;\widetilde{\pi }^i\/_A\/(s)\/-  \nonumber \\
&&\sigma \/_0\/^{AA^{\prime }}\;\widetilde{H}_{AA^{\prime
}}(e)-f_0\/^{AA^{\prime }}\;\widetilde{H}_{AA^{\prime }}(f)-\omega
_0{}^A\/_B\;\widetilde{J}\/_A\/^B-\;a_0\;\widetilde{H}(a)-b_0\;\widetilde{H}%
(b)-  \nonumber \\
&&\psi _0{}^A\/\;\widetilde{S}\/_A\/(q)-\;\varphi _0{}^A\/\;\widetilde{S}%
\/_A\/(s)\/\;-\overline{\psi }_0\/^{A^{\prime }}\;\widetilde{S}_{A^{\prime
}}(\overline{q})-\;\overline{\varphi }_0\/^{A^{\prime }}\;\widetilde{S}%
_{A^{\prime }}(\overline{s}),
\end{eqnarray}
where

$\widetilde{p}^i$\/$_A$\/$^B(\omega )\;=4i\widetilde{\eta }^{ijk}D_j$\ $%
\omega _k{}^B\/_A,$

$\widetilde{p}^i(a)\;=-8\widetilde{\eta }^{ijk}$\ $(2\partial
_jb_k+4f_{jk}+\psi _{jA}\varphi _k$\/$^A),$

$\widetilde{p}^i(b)\;=-8\widetilde{\eta }^{ijk}$\ $(2\partial _ja_k+i\psi
_{jA}\varphi _k$\/$^A),$\ \ \ \ \ \ 

$\widetilde{\pi }^i$\/$_A$\/$(q)$\/ $=-8\widetilde{\eta }^{ijk}\
[iD_j\varphi _{kA}\;+(3a_j-ib_j)\varphi _{kA}\;+2if_{jAA^{\prime }}\overline{%
\psi }_k$\/$^{A^{\prime }}],$\ \ \ \ \ \ \ \ \ 

$\widetilde{\pi }^i$\/$_A$\/$(s)$\/ $=-8\ \widetilde{\eta }^{ijk}[iD_j\psi
_{kA}\;+(3a_j-ib_j)\psi _{kA}\;+2i\sigma _{jAA^{\prime }}\overline{\varphi }%
_k\/^{A^{\prime }}],$%
\begin{equation}  \label{10}
\end{equation}
and

$\widetilde{H}_{AA^{\prime }}(e)\;=64i\widetilde{\eta }^{ijk}f_{iBA^{\prime
}}(f_j\/^{BB^{\prime }}\sigma _{kAB^{\prime }}-\sigma _j\/^{BB^{\prime
}}f_{kAB^{\prime }})+2f_{iAA^{\prime }}\widetilde{p}^i(b)+2\widetilde{\pi }%
^i\/_A$\/$(q)$\/$\overline{\varphi }_{iA^{\prime }},$

$\widetilde{H}_{AA^{\prime }}(f)\;=64i\widetilde{\eta }^{ijk}\sigma
_{iBA^{\prime }}(\sigma _j$\/$^{BB^{\prime }}f_{kAB^{\prime }}$\ $-\;f_j$\/$%
^{BB^{\prime }}\sigma _{kAB^{\prime }})-2\sigma _{iAA^{\prime }}\widetilde{p}%
^i(b)+2\widetilde{\pi }^i\/_A$\/$(s)$\/$\overline{\psi }_{iA^{\prime }},$

$\widetilde{J}_A\/^B\;=D_i\widetilde{p}^i$\/$_A$\/$^B(\omega )\;$

$\widetilde{H}(a)\;=2i[\widetilde{\pi }^i\/_A$\/$(q)\psi _i$\/$^A\;-\;%
\widetilde{\pi }^i\/_A$\/$(s)\varphi _i$\/$^A]$

$\;\;\;\;\;\;\;\;\;\;\;\;\;+16\widetilde{\eta }^{ijk}(2D_if_{jk}+\psi _i$\/$%
^Af_{jAA^{\prime }}$\/$\overline{\psi }_k$\/$^{A^{\prime }}-\varphi
_i\/^A\sigma _{jAA^{\prime }}\/\overline{\varphi }_k\/^{A^{\prime }})\;],$

$\stackrel{\symbol{126}}{H(}b)\;=2[\widetilde{\pi }^i\/_A\/(q)\psi
_i\/^A\;-\;\widetilde{\pi }^i\/_A\/(s)\varphi _i\/^A]-16\ i\widetilde{\eta }%
^{ijk}(\psi _i$\/$^Af_{jAA^{\prime }}$\ $\stackrel{\_}{\psi }_k$\/$%
^{A^{\prime }}-\varphi _i\/^A\sigma _{jAA^{\prime }}\ \stackrel{\_}{\varphi }%
_k\/^{A^{\prime }})\;],$

$\widetilde{S}_A$\/$(q)$\/$\;=-D_i\widetilde{\pi }^i\/_A$\/$(q)-i(3a_i-ib_i)%
\widetilde{\pi }^i\/_A$\/$(q)+\frac 12\varphi _{iA}[\;\widetilde{p}^i(b)+i\;%
\widetilde{p}^i(a)]$

$\;\;\;\;\;\;\;\;\;\;\;\;\;+16\ i\widetilde{\eta }^{ijk}\varphi _{iA}\psi
_{jB}\varphi _k$\/$^B,$

$\widetilde{S}_A$\/$(s)$\/$\;=-D_i\widetilde{\pi }^i\/_A$\/$(s)-i(3a_i-ib_i)%
\widetilde{\pi }^i\/_A$\/\/$(s)-\frac 12\psi _{iA}[\;\widetilde{p}^i(b)+i\;%
\widetilde{p}^i(a)]$

$\;\;\;\;\;\;\;\;\;\;\;\;\;-16\ i\widetilde{\eta }^{ijk}\psi _{iA}\psi
_{jB}\varphi _k$\/$^B,$

$\widetilde{\overline{S}}_{A^{\prime }}(q)=2f_{iA^{\prime }}\/^A\widetilde{%
\pi }^i\/_A$\/$(s),$

$\widetilde{\overline{S}}_{A^{\prime }}(s)=2\sigma _{iA^{\prime }}\/^A%
\widetilde{\pi }^i\/_A$\/$(q).$%
\begin{equation}
\end{equation}
Here we use $\stackrel{\symbol{126}}{\eta }^{ijk}$to denote the Levi-Civita
tensor density on $\Sigma _t\;$and the tilde \symbol{126}over a tensor
density to indicate its weight +1. The meaning of all terms in (35) will be
clear in the following. To pass on to the Hamiltonian formulation we have to
use the Legendre transformation and the Dirac-Bergmann algorithm [17,18].
Calculating the canonical momenta conjugate to all the field variables gives
primary constraints.

$\widetilde{\Phi }^0$\/$_{AA^{\prime }}(e)=\widetilde{p}$\/$^0$\/$%
_{AA^{\prime }}(e)=0,\;\widetilde{\Phi }^i$\/$_{AA^{\prime }}(e)=\widetilde{p%
}^i$\/$_{AA^{\prime }}(e)=0,$

$\widetilde{\Phi }^0$\/$_{AA^{\prime }}(f)=\widetilde{p}^0$\/$_{AA^{\prime
}}(f)=0,\;\widetilde{\Phi }^i$\/$_{AA^{\prime }}(f)=\widetilde{p}^i$\/$%
_{AA^{\prime }}(f)=0,$

$\widetilde{\Phi }^0$\/$_A$\/$^B(\omega )=\widetilde{p}^0$\/$_A$\/$^B(\omega
)=0,\;\widetilde{\Phi }^i$\/$_A$\/$^B(\omega )=\widetilde{p}^i$\/$_A$\/$%
^B(\omega )-4i\stackrel{\symbol{126}}{\eta }^{ijk}R^{+}{}_{ikA}\/^B(\omega
)\approx 0,$

$\widetilde{\Phi }^0$\/$(a)=\widetilde{p}^0$\/$(a)=0,\;\widetilde{\Phi }%
^i(a) $\/$=\widetilde{p}^i(a)$\/$+\stackrel{\symbol{126}}{\eta }%
^{ijk}(16\partial _jb_k+32f_{jk}+8\psi _{jA}\varphi _k\/^A)\approx 0,$

$\widetilde{\Phi }^0(b)$\/$=\widetilde{p}^0(b)$\/$=0,\;\widetilde{\Phi }%
^i(b) $\/$=\widetilde{p}^i(b)$\/$+\stackrel{\symbol{126}}{\eta }%
^{ijk}(16\partial _ja_k+8i\psi _{jA}\varphi _k\/^A)\approx 0,$

$\widetilde{\Phi }^0$\/$_A(q)=\widetilde{\pi }^0$\/$_A(q)=0,$

$\;\widetilde{\Phi }^i$\/$_A(q)=\widetilde{\pi }^i$\/$_A(q)+\stackrel{%
\symbol{126}}{\eta }^{ijk}[8iD_j\varphi _{kA}-8\varphi
_{jA}(3a_k-ib_k)+16f_{jAA^{\prime }}\overline{\psi }_k$\/$^{A^{\prime
}}]\approx 0,$

$\widetilde{\Phi }^0$\/$_A(s)=\widetilde{\pi }^0$\/$_A(s)=0,$

$\;\widetilde{\Phi }^i$\/$_A(s)=\widetilde{\pi }^i$\/$_A(s)+\stackrel{%
\symbol{126}}{\eta }^{ijk}[8iD_j\psi _{kA}-8\psi _{jA}(3a_k-ib_k)+16\sigma
_{jAA^{\prime }}\overline{\varphi }_k$\/$^{A^{\prime }}]\approx 0,$

$\widetilde{\Phi }^0$\/$_{A^{\prime }}(\overline{q})=\widetilde{\pi }^0$\/$%
_{A^{\prime }}(\overline{q})=0,\;\widetilde{\Phi }^i$\/$_{A^{\prime }}(%
\overline{q})=\widetilde{\pi }^i$\/$_{A^{\prime }}(\overline{q})=0,$

$\widetilde{\Phi }^0$\/$_{A^{\prime }}(\overline{s})=\widetilde{\pi }^0$\/$%
_{A^{\prime }}(\overline{s})=0,\;\widetilde{\Phi }^i$\/$_{A^{\prime }}(%
\overline{s})=\widetilde{\pi }^i$\/$_{A^{\prime }}(\overline{s})=0,$%
\begin{equation}
\end{equation}

The basic canonical variables in the theory can then be reduced to$\;\omega
_i{}^A\/_B,\;a_i,\;b_i,\;\;\psi _i$\/$^A,\;\varphi _i$\/$^A$ and their
conjugate momenta$\;\;\widetilde{p}^i$\/$_A$\/$^B(\omega )\;,\;\widetilde{p}%
^i(a),\;\widetilde{p}^i(b),\widetilde{\pi }^i$\/$_A$\/$(q)\;$, and $%
\widetilde{\pi }^i$\/$_A$\/$(s)$. The $\;\omega _i{}^A\/_B\;$is just the
Ashtekar connection.\ The canonical momentum conjugate to $\;\omega
_i{}^A\/_B\;$, however, is not the $\widetilde{\sigma }^i$\/$_A$\/$^B$\ but\
the $\;\widetilde{p}^i$\/$_A$\/$^B(\omega )$ $=4i\widetilde{\eta }%
^{ijk}D_j\omega _k{}^B\/_A$ being different from the Ashtekar theory. The
remaining variables $\sigma \/_0\/^{AA^{\prime }},\;f_0\/^{AA^{\prime
}},\;\;\;\omega _0{}^A\/_B\;,\;a_0,\;b_0$,$\;\psi _0$\/$^A,\varphi _0$\/$^A,%
\overline{\psi }_0$\/$^{A^{\prime }},\;$and\ $\overline{\varphi }_0$\/$%
^{A^{\prime }}\;$play the role of Lagrange multipliers. The $\sigma
\/_i\/^{AA^{\prime }},\;f_i\/^{AA^{\prime }}\;$are neither dynamical
variables nor Lagrange multipliers. The canonical Hamiltonian is

\begin{eqnarray}
H_c &=&\int_{\Sigma _t}\sigma \/_0\/^{AA^{\prime }}\stackrel{}{\widetilde{H}%
_{AA^{\prime }}}(e)+\;f_0\/^{AA^{\prime }}\widetilde{H}_{AA^{\prime
}}(f)+\;\omega _0{}^A\/_B\widetilde{J}\/_A\/^B+a_0\widetilde{H}(a)+b_0%
\widetilde{H}(b)+  \nonumber \\
\; &&\;\;\psi _0{}^A\/\widetilde{S}\/_A\/(q)\/\;+\varphi _0{}^A\/\;%
\widetilde{S}\/_A\/(s)\/+\overline{\psi }_0\/^{A^{\prime }}\widetilde{S}%
_{A^{\prime }}(\stackrel{\_}{q})\;+\overline{\varphi }_0\/^{A^{\prime }}\;%
\widetilde{S}_{A^{\prime }}(\stackrel{\_}{s}).
\end{eqnarray}
Using\ $H_c\;$and the linear combination of the primary constraints with
arbitrary function coefficients we can construct the primary (or total)\
Hamiltonian. Then the consistency conditions {\it i.e. }the requirements of
preserving constraints under time evolution lead to secondary constraints

\begin{eqnarray}
\widetilde{H}_{AA^{\prime }}(e)\; &=&0,\ \widetilde{H}_{AA^{\prime
}}(f)\;=0,\;\widetilde{J}\/_A\/^B\;=0,\;\widetilde{H}(a)\;=0,\;\widetilde{H}%
(b)\;=0,  \nonumber \\
\widetilde{S}\/_A\/(q)\/\; &=&0,\;\widetilde{S}\/_A\/(s)\/\;=0,\;\widetilde{S%
}_{A^{\prime }}(\stackrel{\_}{q})=0,\;\widetilde{S}_{A^{\prime }}(\stackrel{%
\_}{s})=0.\;
\end{eqnarray}
which are the generators of the superconformal group SU(2,2$|$1). In order
to classify the constraints (36) and (38) we have to compute Poisson
brackets between each pairs of them. The complicated results which are given
in the appendix make this classification very difficult. However using Dirac
brackets instead of Poisson brackets one finds that all the constraints are
first class and the constraints (38) are the generators of the
superconformal group SU(2,2$|$1).

In summary, we have given a Hamiltonian formulation of the self-dual
conformal supergravity which is a constrained Hamiltonian system. The
Lagrangian (33) is first order in the time derivatives and the Hamiltonian
(37) results to be a linear combination of the constraints.. This is a
theory of connection dynamics in which one of the basic dynamical variables
is the self-dual spin {}connection ({\it i.e. }the\ Ashtekar connection) $%
\omega _i{}^A\/_B\;$rather than the triad $\sigma _i$\/$^{AB}$ .
Unfortunately, the Dirac bracket structure is very involved in our case ,
and we were not able to compute it explicitly.

\section{Appendix}

In order to classify the constraints we compute the Poisson brackets between
them according to the method given by Casalbuoni [19] the nonvanishing
Poisson brackets are listed here.

The nonvanishing Poisson brackets between the primary constraints are

$\{\widetilde{\Phi }^i$\/$_{AA^{\prime }}(e),\widetilde{\Phi }%
^j(a)\}=32\int_{\Sigma _t}\widetilde{\eta }^{ijk}f_{kAA^{\prime }},$

$\{\widetilde{\Phi }^i$\/$_{AA^{\prime }}(e),\widetilde{\Phi }^j$\/$%
_B(s)\}=16i\int_{\Sigma _t}\widetilde{\eta }^{ijk}\epsilon _{AB}\stackrel{\_%
}{\varphi }_{kA^{\prime }},$

$\{\widetilde{\Phi }^i$\/$_{AA^{\prime }}(f),\widetilde{\Phi }%
^j(a)\}=32\int_{\Sigma _t}\widetilde{\eta }^{ijk}\sigma _{kAA^{\prime }},$

$\{\widetilde{\Phi }^i$\/$_{AA^{\prime }}(f),\widetilde{\Phi }^j$\/$%
_B(q)\}=16i\int_{\Sigma _t}\widetilde{\eta }^{ijk}\epsilon _{AB}\stackrel{\_%
}{\psi }_{kA^{\prime }},$

$\{\;\widetilde{\Phi }^i$\/$_A$\/$^B(\omega ),\widetilde{\Phi }^j$\/$%
_A(q)\}=8i\int_{\Sigma _t}\widetilde{\eta }^{ijk}\delta _C^B\varphi _{kA},$

$\{\;\widetilde{\Phi }^i$\/$_A$\/$^B(\omega ),\widetilde{\Phi }^j$\/$%
_A(s)\}=8i\int_{\Sigma _t}\widetilde{\eta }^{ijk}\delta _C^B\psi _{kA},$

$\{\widetilde{\Phi }^i$\/$_A(q),\;\widetilde{\Phi }^j$\/$_{A^{\prime }}(%
\overline{q})\}=16i\int_{\Sigma _t}\widetilde{\eta }^{ijk}f_{kAA^{\prime }},$

$\{\widetilde{\Phi }^i$\/$_A(s),\;\widetilde{\Phi }^j$\/$_{A^{\prime }}(%
\overline{s})\}=16i\int_{\Sigma _t}\widetilde{\eta }^{ijk}\sigma
_{kAA^{\prime }},$%
\begin{equation}
\end{equation}
The remaining Poisson brackets between the primary constraints vanish. One
can find that the constraints $\widetilde{\Phi }^0$\/$_{AA^{\prime }}(e),\;%
\widetilde{\Phi }^0$\/$_{AA^{\prime }}(f),\;\widetilde{\Phi }^0$\/$_A$\/$%
^B(\omega ),\;\widetilde{\Phi }^0(a),\;\widetilde{\Phi }^0(b),\;\widetilde{%
\Phi }^0$\/$_A(q),\;\widetilde{\Phi }^0$\/$_A(s),\;\widetilde{\Phi }^0$\/$%
_{A^{\prime }}(\overline{q}),\;\widetilde{\Phi }^0$\/$_{A^{\prime }}(%
\overline{s})\;$are first class. In addition there are vanishing Poisson
brackets

$\{\sigma \/_k\/^{AA^{\prime }}\widetilde{\Phi }^i$\/$_{AA^{\prime
}}(e)+f_k\/^{AA^{\prime }}\widetilde{\Phi }^i\/_{AA^{\prime }}(f),\;%
\widetilde{\Phi }^j(a)\}=0,$

$\{\widetilde{\Phi }^i$\/$_{AA^{\prime }}(e)+\widetilde{\Phi }%
^i\/_{AA^{\prime }}(f),\;\overline{\psi }_j\/^{B^{\prime }}\widetilde{\Phi }%
^j\/_B(s)-\overline{\varphi }_j\/^{B^{\prime }}\widetilde{\Phi }%
^j\/_B(q)\}=0,$

$\{\;\widetilde{\Phi }^i$\/$_A$\/$^B(\omega ),\;\psi _j$\/$^C\widetilde{\Phi 
}^j\/_D(q)-\;\varphi _j$\/$^C\widetilde{\Phi }^j\/_D(s)\}=0,$

$\{\widetilde{\Phi }^i$\/$_A(q)+\widetilde{\Phi }^i\/_A(s),\;\widetilde{\Phi 
}^j$\/$_{A^{\prime }}(\overline{q})\sigma \/_k\/^{BA^{\prime }}+\widetilde{%
\Phi }^j$\/$_{A^{\prime }}(\overline{s})f\/_k\/^{BA^{\prime }}\}=0.$%
\begin{equation}
\end{equation}
This means that there are more primary constraints which are first class.

The nonvanishing Poisson brackets between the primary constraints and the
secondary constraints are

$\{\widetilde{\Phi }^i$\/$_{AA^{\prime }}(e),\widetilde{H}_{BB^{\prime
}}(e)\;\}=-64i\int_{\Sigma _t}\widetilde{\eta }^{ijk}(f_{jAB^{\prime
}}f_{kBA^{\prime }}+f_j$\/\/$^C$\/$_{A^{\prime }}f_{kCB^{\prime }}\epsilon
_{AB}),$

$\{\widetilde{\Phi }^i$\/$_{AA^{\prime }}(e),\widetilde{H}_{BB^{\prime
}}(f)\;\}=$

$\;\;\;\int_{\Sigma _t}2\epsilon _{AB}\epsilon _{A^{\prime }B^{\prime }}%
\widetilde{p}^i(b)+64i\widetilde{\eta }^{ijk}\epsilon _{A^{\prime }B^{\prime
}}(\sigma _{jA}$\/$^{C^{\prime }}f_{kBC^{\prime }}-f_{jA}$\/$^{C^{\prime
}}\sigma _{kBC^{\prime }})+$

\ \ \ \ \ \ \ $64i\widetilde{\eta }^{ijk}(\sigma _{jAB^{\prime
}}f_{kBA^{\prime }}+f_J$\/$^C$\/$_{A^{\prime }}\sigma _{kCB^{\prime }}$\ $%
\epsilon _{AB}),$

$\{\widetilde{\Phi }^i$\/$_{AA^{\prime }}(e),\widetilde{H}(a)\;\}=$\ $%
\int_{\Sigma _t}\widetilde{\eta }^{ijk}(32D_jf_{kAA^{\prime }}-16\varphi
_{jA}\overline{\varphi }_{kA^{\prime }}),$

$\{\widetilde{\Phi }^i$\/$_{AA^{\prime }}(e),\widetilde{H}(b)\;\}=16i$\ $%
\int_{\Sigma _t}\widetilde{\eta }^{ijk}\varphi _{jA}\overline{\varphi }%
_{kA^{\prime }},$\ \ \ 

$\{\widetilde{\Phi }^i$\/$_{AA^{\prime }}(f),\widetilde{H}_{BB^{\prime
}}(e)\;\}=$

$\;\;\;-\int_{\Sigma _t}2\epsilon _{AB}\epsilon _{A^{\prime }B^{\prime }}%
\widetilde{p}^i(b)+64i\widetilde{\eta }^{ijk}\epsilon _{A^{\prime }B^{\prime
}}(\sigma _{jA}$\/$^{C^{\prime }}f_{kBC^{\prime }}-f_{jA}$\/$^{C^{\prime
}}\sigma _{kBC^{\prime }})+$

\ \ \ \ \ \ \ $64i\widetilde{\eta }^{ijk}(\sigma _{jAB^{\prime
}}f_{kBA^{\prime }}+f_j$\/$^C$\/$_{A^{\prime }}\sigma _{kCB^{\prime }}$\ $%
\epsilon _{AB}),$

$\{\widetilde{\Phi }^i$\/$_{AA^{\prime }}(f),\widetilde{H}_{BB^{\prime
}}(f)\;\}=-64i\int_{\Sigma _t}\widetilde{\eta }^{ijk}(\sigma _{jAB^{\prime
}}\sigma _{kBA^{\prime }}-\sigma _j$\/\/$^C$\/$_{A^{\prime }}\sigma
_{kCB^{\prime }}\epsilon _{AB}),$

$\{\widetilde{\Phi }^i$\/$_{AA^{\prime }}(f),\widetilde{H}(a)\;\}=$\ $%
\int_{\Sigma _t}\widetilde{\eta }^{ijk}(32D_j\sigma _{kAA^{\prime }}+16\psi
_{jA}\overline{\psi }_{kA^{\prime }}),$

$\{\widetilde{\Phi }^i$\/$_{AA^{\prime }}(f),\widetilde{H}(b)\;\}=-16i$\ $%
\int_{\Sigma _t}\widetilde{\eta }^{ijk}\psi _{jA}\overline{\psi }%
_{kA^{\prime }},$

$\{\;\widetilde{\Phi }^i$\/$_A$\/$^B(\omega ),J$\/$_C$\/$^D\}=8i\int_{\Sigma
_t}\widetilde{\eta }^{ijk}(\delta _A^D\omega $\/$_{jE}$\/$^B\omega $\/$_{kC}$%
\/$^E+\delta _C^B\omega $\/$_{jA}$\/$^E\omega $\/$_{kE}$\/$^D)+$

\ \ \ \ \ \ \ \ \ \ \ \ \ \ \ \ \ \ \ \ \ \ \ $\int_{\Sigma _t}$\ $\delta
_A^D$\ $\stackrel{\symbol{126}}{p}^i$\/$_C$\/$^B(\omega )-\delta _C^B%
\stackrel{\symbol{126}}{p}^i$\/$_A$\/$^D(\omega ),$

$\{\widetilde{\Phi }^i$\/$_A$\/$^B(\omega ),\widetilde{S}$\/$_C$\/$(q)$\/$%
\}= $\ $\int_{\Sigma _t}\delta _C^B\widetilde{\pi }^i$\/$_A(q),$

$\{\widetilde{\Phi }^i$\/$_A$\/$^B(\omega ),\stackrel{\symbol{126}}{S}$\/$_C$%
\/$(s)$\/$\}=$\ $\int_{\Sigma _t}\delta _C^B\widetilde{\pi }^i$\/$_A(s),$

$\{\widetilde{\Phi }^i(a),\widetilde{S}$\/$_A$\/$(q)$\/$\}=\int_{\Sigma _t}8%
\widetilde{\eta }^{ijk}[\omega $\/$_{jA}$\/$^B\varphi
_{kB}+(3ia_j+b_j)\varphi _{kA}]+3i\widetilde{\pi }^i$\/$_A(q),$

$\{\widetilde{\Phi }^i(a),\widetilde{S}$\/$_A$\/$(s)$\/$\}=\int_{\Sigma _t}-8%
\widetilde{\eta }^{ijk}[\omega $\/$_{jA}$\/$^B\psi _{kB}+(3ia_j+b_j)\psi
_{kA}]+3i\widetilde{\pi }^i$\/$_A(s),$

$\{\widetilde{\Phi }^i(b),\widetilde{S}$\/$_A$\/$(q)$\/$\}=\int_{\Sigma _t}8%
\widetilde{\eta }^{ijk}[\omega $\/$_{jA}$\/$^B\varphi
_{kB}-(3ia_j+b_j)\varphi _{kA}]+\widetilde{\pi }^i$\/$_A(q),$

$\{\widetilde{\Phi }^i(b),\widetilde{S}$\/$_A$\/$(s)$\/$\}=\int_{\Sigma _t}-8%
\widetilde{\eta }^{ijk}[\omega $\/$_{jA}$\/$^B\psi _{kB}-(3ia_j+b_j)\psi
_{kA}]+\widetilde{\pi }^i$\/$_A(s),$

$\{\widetilde{\Phi }^i$\/$_A(q),\widetilde{H}_{BB^{\prime
}}(e)\}=16i\int_{\Sigma _t}\widetilde{\eta }^{ijk}\varphi
_{jA}f_{kBB^{\prime }},$

$\{\widetilde{\Phi }^i$\/$_A(q),\widetilde{H}_{BB^{\prime
}}(f)\}=-16i\int_{\Sigma _t}\widetilde{\eta }^{ijk}\varphi _{jA}\sigma
_{kBB^{\prime }},$

$\{\widetilde{\Phi }^i$\/$_C(q),\widetilde{J}_A\/^B\}=8i\int_{\Sigma _t}%
\widetilde{\eta }^{ijk}(\delta _C^B\omega \/_{jA}\/^D\varphi _{kD}-\omega _j$%
\/$^B$\/$_C\varphi _{kA}),$

$\{\widetilde{\Phi }^i$\/$_A(q),\widetilde{H}(a)\}=\int_{\Sigma _t}16%
\widetilde{\eta }^{ijk}[-\omega _{jAB}\varphi _k\/^B\;+(3ia_j+b_j)\varphi
_{kA}-$

\ \ \ \ \ \ \ \ \ \ \ \ \ \ \ \ \ \ \ \ \ \ \ \ \ \ $\;\;\;\;\;f_{jAA^{%
\prime }}\overline{\psi }_k$\/$^{A^{\prime }}]+2i\widetilde{\pi }^i$\/$%
_A(q), $

$\{\widetilde{\Phi }^i$\/$_A(q),\widetilde{H}(b)\}=\int_{\Sigma _t}16%
\widetilde{\eta }^{ijk}[-i\omega _{jAB}\varphi _k\/^B\;+(3a_j-ib_j)\varphi
_{kA}+$

\ \ \ \ \ \ \ \ \ \ \ \ \ \ \ \ \ \ \ \ \ \ \ \ \ $\;\;\;\;\;\;if_{jAA^{%
\prime }}\overline{\psi }_k$\/$^{A^{\prime }}]+2i\widetilde{\pi }^i$\/$%
_A(q), $

$\{\widetilde{\Phi }^i$\/$_A(q),\widetilde{S}$\/$_B$\/$(q)$\/$%
\}=8i\int_{\Sigma _t}\widetilde{\eta }^{ijk}\varphi _{jA}\varphi _{kB},$

$\{\widetilde{\Phi }^i$\/$_A(q),\widetilde{S}$\/$_B$\/$(s)$\/$%
\}=-\int_{\Sigma _t}8i\widetilde{\eta }^{ijk}[\omega $\/$_{jAC}$\/$\omega $\/%
$_{kB}$\/$^C+2\omega $\/$_{jAB}$\/$(3ia_k+b_k)-\varphi _{jA}\psi _{kB}+$

\ \ \ \ \ \ \ \ \ \ \ \ \ \ \ \ \ \ \ \ \ \ \ \ \ $2\epsilon _{AB}\varphi
_{jC}\psi _k\/^C]+\frac 12\epsilon _{AB}[i\widetilde{p}^i(a)+\;\widetilde{p}%
^i(b)],$

$\{\widetilde{\Phi }^i$\/$_A(s),\widetilde{H}_{BB^{\prime
}}(e)\}=-16i\int_{\Sigma _t}\widetilde{\eta }^{ijk}\psi _{jA}f_{kBB^{\prime
}},$

$\{\widetilde{\Phi }^i$\/$_A(s),\widetilde{H}_{BB^{\prime
}}(f)\}=16i\int_{\Sigma _t}\widetilde{\eta }^{ijk}\varphi _{jA}\sigma
_{kBB^{\prime }},$

$\{\widetilde{\Phi }^i$\/$_C(s),\widetilde{J}_A\/^B\}=8i\int_{\Sigma _t}%
\widetilde{\eta }^{ijk}(\delta _C^B\omega \/_{jA}\/^D\psi _{kD}-\omega _j$\/$%
^B$\/$_C\psi _{kA}),$

$\{\widetilde{\Phi }^i$\/$_A(s),\widetilde{H}(a)\}=\int_{\Sigma _t}16%
\widetilde{\eta }^{ijk}[-\omega _{jAB}\psi _k\/^B\;+(3ia_j+b_j)\psi _{kA}+$

\ \ \ \ \ \ \ \ \ \ \ \ \ \ \ \ \ \ \ \ \ \ \ \ \ \ $\;\;\;\;\sigma
_{jAA^{\prime }}\overline{\varphi }_k$\/$^{A^{\prime }}]-2i\widetilde{\pi }%
^i $\/$_A(s),$

$\{\widetilde{\Phi }^i$\/$_A(s),\widetilde{H}(b)\}=\int_{\Sigma _t}16%
\widetilde{\eta }^{ijk}[i\omega _{jAB}\psi _k\/^B\;+(3a_j-ib_j)\psi _{kA}-$

\ \ \ \ \ \ \ \ \ \ \ \ \ \ \ \ \ \ \ \ \ \ \ \ \ $\;\;\;\;\;\;i\sigma
_{jAA^{\prime }}\overline{\varphi }_k$\/$^{A^{\prime }}]-2i\widetilde{\pi }%
^i $\/$_A(s),$

$\{\widetilde{\Phi }^i$\/$_A(s),\widetilde{S}$\/$_B$\/$(q)$\/$%
\}=-\int_{\Sigma _t}8i\widetilde{\eta }^{ijk}[\omega $\/$_{jAC}$\/$\omega $\/%
$_{kB}$\/$^C+2\omega $\/$_{jAB}$\/$(3ia_k+b_k)+\psi _{jA}\varphi _{kB}-$

\ \ \ \ \ \ \ \ \ \ \ \ \ \ \ \ \ \ \ \ \ \ \ \ \ $2\epsilon _{AB}\varphi
_{jC}\psi _k\/^C]+\frac 12\epsilon _{AB}[i\widetilde{p}^i(a)+\;\widetilde{p}%
^i(b)],$

$\{\widetilde{\Phi }^i$\/$_A(s),\widetilde{S}$\/$_B$\/$(s)$\/$%
\}=8i\int_{\Sigma _t}\widetilde{\eta }^{ijk}\psi _{jA}\psi _{kB},$

$\{\widetilde{\Phi }^i$\/$_{A^{\prime }}(\overline{q}),\widetilde{H}%
_{BB^{\prime }}(f)\}=-2\int_{\Sigma _t}\widetilde{\pi }^i$\/$_B(s)\epsilon
_{A^{\prime }B^{\prime }},$

$\{\widetilde{\Phi }^i$\/$_{A^{\prime }}(\overline{s}),\widetilde{H}%
_{BB^{\prime }}(e)\}=-2\int_{\Sigma _t}\widetilde{\pi }^i$\/$_B(q)\epsilon
_{A^{\prime }B^{\prime }}.$%
\begin{equation}
\end{equation}

The nonvanishing Poisson brackets between the secondary constraints are

$\{\widetilde{H}_{AA^{\prime }}(e),\widetilde{S}\/_B\/(q)\}=2\int_{\Sigma
_t}f_{iAA^{\prime }}\widetilde{\pi }^i$\/$_B(q),$

$\{\widetilde{H}_{AA^{\prime }}(e),\widetilde{S}\/_B\/(s)\}=2\int_{\Sigma
_t}f_{iAA^{\prime }}\widetilde{\pi }^i$\/$_B(s),$

$\{\widetilde{H}_{AA^{\prime }}(f),\widetilde{S}\/_B\/(q)\}=-2\int_{\Sigma
_t}\sigma _{iAA^{\prime }}\widetilde{\pi }^i$\/$_B(q),$

$\{\widetilde{H}_{AA^{\prime }}(f),\widetilde{S}\/_B\/(s)\}=2\int_{\Sigma
_t}\sigma _{iAA^{\prime }}\widetilde{\pi }^i$\/$_B(s),$

$\{\widetilde{J}_A\/^B,\widetilde{S}$\/$_C$\/$(q)$\/$\}=\int_{\Sigma
_t}\delta _C^B\omega \/_{iA}\/^D\widetilde{\pi }^i$\/$_D(q)-\omega
\/_{iC}\/^B\widetilde{\pi }^i$\/$_A(q),$

$\{\widetilde{J}_A\/^B,\widetilde{S}$\/$_C$\/$(s)$\/$\}=\int_{\Sigma
_t}\delta _C^B\omega \/_{iA}\/^D\widetilde{\pi }^i$\/$_D(s)-\omega
\/_{iC}\/^B\widetilde{\pi }^i$\/$_A(s),$

$\{\widetilde{H}(a),\widetilde{S}\/_A\/(q)\}=\int_{\Sigma _t}2i\omega
\/_{iA}\/^B\widetilde{\pi }^i$\/$_B(q)-2(3a_i-ib_i)\widetilde{\pi }%
^i\/_A(q)-[\widetilde{p}^i(a)-i\;\widetilde{p}^i(b)]\varphi _{iA}-$

\ \ \ \ \ \ \ \ \ \ \ \ \ \ \ \ \ \ \ \ \ \ $16\widetilde{\eta }%
^{ijk}[\omega \/_{iA}\/^Bf_{jBB^{\prime }}+(3ia_i+b_i)f_{jAB^{\prime }}%
\overline{\psi }_k$\/$^{B^{\prime }}+32\widetilde{\eta }^{ijk}\varphi
_{iA}\psi _{jB}\varphi _k\/^B,$

$\{\widetilde{H}(a),\widetilde{S}\/_A\/(s)\}=\int_{\Sigma _t}-2i\omega
\/_{iA}\/^B\widetilde{\pi }^i$\/$_B(s)+2(3a_i-ib_i)\widetilde{\pi }%
^i\/_A(s)-[\widetilde{p}^i(a)-i\;\widetilde{p}^i(b)]\psi _{iA}-$

\ \ \ \ \ \ \ \ \ \ \ \ \ \ \ \ \ \ \ \ \ \ $16\widetilde{\eta }%
^{ijk}[\omega \/_{iA}\/^B\sigma _{jBB^{\prime }}+(3ia_i+b_i)\sigma
_{jAB^{\prime }}\overline{\varphi }_k$\/$^{B^{\prime }}-96\widetilde{\eta }%
^{ijk}\psi _{iA}\psi _{jB}\varphi _k\/^B,$

$\{\widetilde{H}(a),\widetilde{\overline{S}}_{A^{\prime }}(\overline{q)}%
\}=\int_{\Sigma _t}4if_i$\/$^A$\/$_{A^{\prime }}\widetilde{\pi }^i$\/$%
_A(s)-32\widetilde{\eta }^{ijk}f_i$\/$^A$\/$_{A^{\prime }}\sigma
_{jAB^{\prime }}\overline{\varphi }_k$\/$^{B^{\prime }},$

$\{\widetilde{H}(a),\widetilde{\overline{S}}_{A^{\prime }}(\overline{s)}%
\}=\int_{\Sigma _t}-4i\sigma _i$\/$^A$\/$_{A^{\prime }}\widetilde{\pi }^i$\/$%
_A(q)+32\widetilde{\eta }^{ijk}\sigma _i$\/$^A$\/$_{A^{\prime
}}f_{jAB^{\prime }}\overline{\psi }_k$\/$^{B^{\prime }},$

$\{\widetilde{H}(b),\widetilde{S}\/_A\/(q)\}=\int_{\Sigma _t}2\omega
\/_{iA}\/^B\widetilde{\pi }^i$\/$_B(q)+2(3ia_i+b_i)\widetilde{\pi }%
^i\/_A(q)+[i\widetilde{p}^i(a)+\;\widetilde{p}^i(b)]\varphi _{iA}-$

\ \ \ \ \ \ \ \ \ \ \ \ \ \ \ \ \ \ \ \ \ \ $16\widetilde{\eta }%
^{ijk}[i\omega \/_{iA}\/^Bf_{jBB^{\prime }}-(3a_i-ib_i)f_{jAB^{\prime }}%
\overline{\psi }_k$\/$^{B^{\prime }}+32\widetilde{\eta }^{ijk}\varphi
_{iA}\psi _{jB}\varphi _k\/^B,$

$\{\widetilde{H}(b),\widetilde{S}\/_A\/(s)\}=\int_{\Sigma _t}-2\omega
\/_{iA}\/^B\widetilde{\pi }^i$\/$_B(s)-2(3ia_i+b_i)\widetilde{\pi }%
^i\/_A(s)+[i\widetilde{p}^i(a)+\;\widetilde{p}^i(b)]\psi _{iA}-$

\ \ \ \ \ \ \ \ \ \ \ \ \ \ \ \ \ \ \ \ \ \ $16\widetilde{\eta }%
^{ijk}[i\omega \/_{iA}\/^B\sigma _{jBB^{\prime }}-(3a_i-ib_i)\sigma
_{jAB^{\prime }}\overline{\varphi }_k$\/$^{B^{\prime }}+96\widetilde{\eta }%
^{ijk}\psi _{iA}\psi _{jB}\varphi _k\/^B,$

$\{\widetilde{H}(b),\widetilde{\overline{S}}_{A^{\prime }}(\overline{q)}%
\}=\int_{\Sigma _t}4f_i$\/$^A$\/$_{A^{\prime }}\widetilde{\pi }^i$\/$%
_A(s)+32i\widetilde{\eta }^{ijk}f_i$\/$^A$\/$_{A^{\prime }}\sigma
_{jAB^{\prime }}\overline{\varphi }_k$\/$^{B^{\prime }},$

$\{\widetilde{H}(b),\widetilde{\overline{S}}_{A^{\prime }}(\overline{s)}%
\}=\int_{\Sigma _t}-4\sigma _i$\/$^A$\/$_{A^{\prime }}\widetilde{\pi }^i$\/$%
_A(q)-32i\widetilde{\eta }^{ijk}\sigma _i$\/$^A$\/$_{A^{\prime
}}f_{jAB^{\prime }}\overline{\psi }_k$\/$^{B^{\prime }},$

$\{\widetilde{S}\/_A\/(q),\widetilde{S}\/_B\/(s)\}=\int_{\Sigma _t}-%
\widetilde{\pi }^i$\/$_A(q)\psi _{iB}+\widetilde{\pi }^i$\/$_A(s)\varphi
_{iB}+\epsilon _{AB}(3a_i-ib_i)[\widetilde{p}^i(a)-i\;\widetilde{p}^i(b)]+$

\ \ \ \ \ \ \ \ \ \ \ \ \ \ \ \ \ \ \ \ \ \ \ $16i\widetilde{\eta }%
^{ijk}(\omega _{iAC}\psi _{jB}\varphi _k\/^C-\omega _{iBC}\varphi _{jA}\psi
_k\/^C)+$

$\;\;\;\;\;\;\;\;\;\;\;\;\;\;\;\;\;\;\;\;\;\;\;32\epsilon _{AB}\widetilde{%
\eta }^{ijk}(3a_i-ib_i)\psi _{jC}\varphi _k\/^C,$

$\{\widetilde{S}\/_A\/(q),\widetilde{\overline{S}}_{A^{\prime }}(\overline{q)%
}\}=\int_{\Sigma _t}f_{iAA^{\prime }}[i\widetilde{p}^i(a)+\;\widetilde{p}%
^i(b)]+32i\widetilde{\eta }^{ijk}(f_{iAA^{\prime }}\psi _{jB}\varphi
_k\/^B+f_{iBA^{\prime }}\varphi _{jA}\psi _k\/^B),$

$\{\widetilde{S}\/_A\/(q),\widetilde{\overline{S}}_{A^{\prime }}(\overline{s)%
}\}=32i\int_{\Sigma _t}\widetilde{\eta }^{ijk}\sigma _i$\/$^B$\/$_{A^{\prime
}}\varphi _{jA}\varphi _{kB},$

$\{\widetilde{S}\/_A\/(s),\widetilde{\overline{S}}_{A^{\prime }}(\overline{q)%
}\}=32i\int_{\Sigma _t}\widetilde{\eta }^{ijk}f_i$\/$^B$\/$_{A^{\prime
}}\psi _{jA}\psi _{kB},$

$\{\widetilde{S}\/_A\/(s),\widetilde{\overline{S}}_{A^{\prime }}(\overline{s)%
}\}=-\int_{\Sigma _t}\sigma _{iAA^{\prime }}[i\widetilde{p}^i(a)+\;%
\widetilde{p}^i(b)]+32i\widetilde{\eta }^{ijk}(\sigma _{iAA^{\prime }}\psi
_{jB}\varphi _k\/^B+\sigma _{iBA^{\prime }}\varphi _{jA}\psi _k\/^B).$%
\begin{equation}
\end{equation}
It is very difficult to classify constraints using these Poisson brackets..
Only two first class secondary constraints can be found out:

$\sigma \/_i\/^{AA^{\prime }}\widetilde{H}_{AA^{\prime
}}(e)+f_i\/^{AA^{\prime }}\widetilde{H}_{AA^{\prime }}(f)\;$and $\widetilde{H%
}(a)-i\widetilde{H}(a).$

\end{document}